\documentclass[pra,twocolumn,superscriptaddress,10pt]{revtex4-1}
\usepackage{times,amssymb,amsfonts,amsmath}
\usepackage[dvipsnames]{xcolor}
\usepackage{mathrsfs}
\usepackage{bm}
\newtheorem{theorem}{Theorem}

\newtheorem{corollary}{Corollary}

\begin{document}

\title{Five-Dimensional Compatibility and Stratified Local-Unitary Structure of Pure Three-Qubit States}

\author{Wei Song}\email{wsong1@mail.ustc.edu.cn}
\affiliation{School of Physics and Materials Engineering, Hefei Normal University,
Hefei 230601, China}

\author{Xiao-Lan Zong}\email{zxl@hfnu.edu.cn}
\affiliation{School of Physics and Materials Engineering, Hefei Normal University,
Hefei 230601, China}

\author{Ming Yang}
\affiliation{School of Physics, Anhui University, Hefei, 230601, China}
\affiliation{Leibniz International Joint Research Center of Materials Sciences of Anhui Province, Anhui University, Hefei 230601, China}

\begin{abstract}
We investigates the joint compatibility problem of the three pairwise concurrences, the three-tangle, and
the Kempe invariant in arbitrary pure three-qubit states, and gives necessary and sufficient conditions for
these five quantities to be simultaneously realizable by the same pure state. This five-dimensional compatibility region can be viewed as a fiber structure over the previously derived four-dimensional reachable region. We further find that the continuous local-unitary completeness of the four tangle coordinates depends on where the state lies in the allowed region. In the generic interior, one continuous degree of freedom remains. It is fixed if the three-tangle vanishes or a pairwise concurrence becomes zero, and also at the boundary of the four-dimensional tangle region. The Kempe invariant gives the simplest algebraic parametrization of the fifth coordinate, but the same degree of freedom may be described by a known pure-state entanglement monotone. The compatibility conditions can also be expressed in terms of concurrence-based bipartite measures and multipartite quantities fixed by the pairwise concurrences and the three-tangle.

\end{abstract}

\maketitle

\section{Introduction}

Quantifying entanglement has always been a central problem in quantum information theory, and various entanglement measures have been developed to describe different quantum correlations in bipartite and multipartite systems\cite{Horodecki2009,Wootters1998}. However, in multipartite systems, merely knowing the magnitudes of various entanglements does not completely characterize the structure of a quantum state, because different entanglement measures cannot vary independently of each other. This gives rise to a more fundamental question than quantifying entanglement: in what ways can different forms of entanglement coexist within a single quantum state?

Three-qubit systems are the simplest nontrivial system for studying this problem. For a pure three-qubit state, bipartite entanglement can coexist in the three bipartite reduced states, while the whole system can also possess genuine tripartite quantum correlations\cite{Dur2000,Verstraete2002}. The CKW monogamy relation and its generalizations\cite{Coffman2000,Osborne2006,Bai2014,GuoGour2019,KoashiWinter2004,Adesso2006,Gour2017,Fei2018,Song2016,Dong2026,Camalet2017} characterize the constraints on entanglement distribution among different bipartitions. On the other hand, geometric methods such as the entanglement polytope\cite{Walter2013}, starting from the SLOCC classification, describe the allowed range of the spectra of local reduced states. We focuses on a direction complementary to these issues: given several entanglement measures and local unitary invariants, can they be simultaneously realized by the same pure three-qubit state?

For any pure three-qubit state, Allen, Bucicovschi, and Meyer~\cite{Allen2017} have derived an exact compatibility relation among the three bipartite tangles and the three-tangle. However, these four quantities are not sufficient to fully characterize the continuous local-unitary structure of a pure three-qubit state. A generic pure three-qubit state has five continuous parameters up to local unitaries. Even after fixing the three pairwise concurrences and the three-tangle, one continuous degree of freedom remains generically. The Kempe invariant provides a natural coordinate for this remaining degree of freedom~\cite{Kempe1999,Sudbery2001,Acin2000}.

In this paper, based on the three bipartite concurrences and the three-tangle, we further add the Kempe invariant and provide necessary and sufficient conditions for these five coordinates to be simultaneously realized by the same pure three-qubit state. In the permutation-symmetric sector, our conditions reduce to those of Ref.~\cite{MeillMeyer2017}. For a generic pure three-qubit state, fixing the three pairwise concurrences and the three-tangle still leaves one continuous parameter free. The value is fixed on several lower-dimensional strata, such as the zero-three-tangle sector, states with a vanishing pairwise concurrence, and the boundary of the compatibility region. Moreover, we adopt the Kempe invariant mainly because it is a convenient coordinate for describing this direction; this degree of freedom can also be parametrized by the pure-state entanglement monotone constructed by Oreshkov and Brun~\cite{OreshkovBrun2006}. The corresponding compatibility conditions can also be reformulated to apply to bipartite entanglement measures based on concurrence, as well as to multipartite entanglement measures determined by the pairwise concurrences and the three-tangle coordinates.

\section{Five-Dimensional Compatibility Body}

We define
\begin{equation}
x=C_{AB}^{2},y=C_{AC}^{2},z=C_{BC}^{2},t=\tau_3,
\label{eq:xyztdef}
\end{equation}
and adopt the Kempe invariant in the following form:
\begin{equation}
\kappa = 3\operatorname{Tr}[(\rho_A\otimes\rho_B)\rho_{AB}]
-\operatorname{Tr}\rho_A^3 -\operatorname{Tr}\rho_B^3.
\label{eq:kempe_def}
\end{equation}
For later use, define
\begin{equation}
S=x+y+z,Q=xy+xz+yz,p=\sqrt{xyz},
\label{eq:SQpdef}
\end{equation}
and
\begin{equation}
q=\sqrt{(t+x)(t+y)(t+z)}.
\label{eq:qdef}
\end{equation}
In Ref.~\cite{Allen2017}, the reachable region of pure three-qubit states is given by:
\begin{equation}
\Omega_4 = \left\{ (x,y,z,t): 0\le x,y,z,t\le1,\quad F_-(x,y,z,t)\le0 \right\},
\label{eq:Omega4}
\end{equation}
where
\begin{equation}
F_-(x,y,z,t) = t^2+(S-1)t+Q-2p.
\label{eq:allen_base}
\end{equation}
When $(x,y,z,t)$ is fixed, one can separate out the part of the Kempe invariant that is directly determined by these four entanglement coordinates. To this end, we define the reduced Kempe coordinate
\begin{equation}
\mu = \frac43(\kappa-1)+S+t,
\label{eq:mu_def}
\end{equation}
or equivalently,
\begin{equation}
\kappa = 1-\frac34(S+t)+\frac34\mu.
\label{eq:kappa_from_mu}
\end{equation}
With this coordinate, the five-dimensional compatibility condition can be stated as follows.

\begin{theorem}[Five-dimensional compatibility].
\label{thm:five_body}
A set of five coordinates $(x,y,z,t,\kappa)$ can be realized by some normalized pure three-qubit state if and only if
\begin{equation}
0\le x,y,z,t\le1, \qquad F_-(x,y,z,t)\le0,
\label{eq:five_base_condition}
\end{equation}
and
\begin{equation}
\mu^2\le xyz, \qquad \mu+t\ge \sqrt{(t+x)(t+y)(t+z)},
\label{eq:five_mu_condition}
\end{equation}
where $\mu$ is defined as above.
\end{theorem}

Therefore, the five-dimensional compatibility problem amounts to determining two conditions that the remaining continuous coordinate must satisfy on the basis of the known four-dimensional reachable region. In Eq.~\eqref{eq:five_mu_condition}, the first inequality corresponds to the phase feasibility in the Ac{\'i}n canonical form, and the second inequality ensures that the remaining standard parameters can take real values. The detailed derivation is provided in Appendix~\ref{app:kempe_acin}; the proof of Theorem~\ref{thm:five_body} is given in Appendix~\ref{app:five_body_proof}, while explicit reconstruction formulas and representative states for the degenerate sectors are collected in Appendix~\ref{app:constructive}.

For any fixed four-dimensional point in $\Omega_4$, the allowed values of the reduced Kempe coordinate form an interval
\begin{equation}
\max\{-p,\, q-t\}\le \mu \le p.
\label{eq:mu_interval}
\end{equation}
The five-dimensional reachable region is obtained by allowing the Kempe coordinate to vary over its admissible interval above each point of the four-dimensional compatibility region. In general, this interval has nonzero length; on special strata it collapses to a single point. Equation~\eqref{eq:five_mu_condition} gives the complete compatibility conditions for the five coordinates. At fixed $(x,y,z,t)$, they reduce to the allowed interval for $\mu$ in Eq.~\eqref{eq:mu_interval}.

The five-dimensional reachable region can be written as
\begin{equation}
\Omega_5 = \left\{ (x,y,z,t,\kappa): (x,y,z,t)\in\Omega_4,\; \mu\in I_\mu \right\},
\label{eq:interval_fibration}
\end{equation}
where $I_\mu=[\max\{-p,q-t\},p]$. Projecting out the Kempe coordinate recovers the four-dimensional compatibility region found by Allen--Bucicovschi--Meyer. For fixed $(x,y,z,t)$, the interval $I_\mu$ describes the remaining continuous local-unitary degree of freedom.

For permutation-symmetric pure three-qubit states, the three pairwise concurrences are equal, so we have
$x=y=z=C^2$. In this case, Eq.~\eqref{eq:five_mu_condition} reduces to
\begin{equation}
\mu^2\le C^6, \qquad \mu+t\ge(C^2+t)^{3/2},
\label{eq:symmetric_mu}
\end{equation}
that is,
\begin{equation}
\max\{-C^3,(C^2+t)^{3/2}-t\} \le\mu\le C^3.
\end{equation}
 With the three-tangle normalization written in the convention of Ref.~\cite{MeillMeyer2017}, the above relations, expressed in terms of $\kappa$, reproduce the Meill--Meyer boundary for permutation-symmetric pure three-qubit states. Therefore, under the symmetric condition $x=y=z$, the general five-dimensional compatibility region reduces to the symmetric-state reachable region found by Meill and Meyer.

\section{Stratified Completeness of Tangle-Based Entanglement Descriptions}
\label{sec:stratified_completeness}
For any point $b=(x,y,z,t)\in\Omega_4$ in the four-dimensional reachable region, all allowed values of the reduced Kempe coordinate $\mu$ constitute
\begin{equation}
\mathcal F_b
=
\{\mu:\mu^2\le xyz,\ \mu+t\ge q\}
=
[\max\{-p,q-t\},p].
\label{eq:fiber_def}
\end{equation}
This fibre may be a closed interval of nonzero length or may collapse to a single point. We denote its continuous dimension by
$\nu(b)=\dim_{\rm cont}\mathcal F_b$: when $\mathcal F_b$ is a nondegenerate interval, $\nu(b)=1$; when it degenerates to a point, $\nu(b)=0$. The interval form of $\mathcal F_b$ then gives the following classification.

\begin{corollary}[Stratified continuous-LU completeness]
\label{cor:stratified_completeness}
For any $b=(x,y,z,t)\in\Omega_4$, we have
\begin{equation}
\nu(b) =
\begin{cases}
1, & txyz>0 \;\text{and}\; F_-(x,y,z,t)<0,\\[1mm]
0, & txyz=0 \;\text{or}\; F_-(x,y,z,t)=0.
\end{cases}
\label{eq:fiber_dimension}
\end{equation}
In the interior of the four-dimensional reachable region, the four tangles generically leave one continuous local-unitary parameter undetermined. On the degenerate strata and on the boundary, the reduced Kempe coordinate is fixed by $(x,y,z,t)$.
\end{corollary}

Equation~\eqref{eq:fiber_dimension} follows directly from $I_\mu=[\max\{-p,q-t\},p]$. The fibre has nonzero length if and only if
\begin{equation}
p>\max\{-p,q-t\},
\end{equation}
which is equivalent to simultaneously satisfying $p>0,\qquad t+p>q$. Since $p=\sqrt{xyz}$, the first condition is equivalent to $xyz>0$. On the other hand, for any physical point in $\Omega_4$ we have
\begin{equation}
(t+p)^2-q^2=-tF_-.
\label{eq:factor_relation_main}
\end{equation}
Thus, $t+p>q$ holds if and only if $t>0$ and $F_-<0$. This yields Eq.~\eqref{eq:fiber_dimension}. For a more detailed derivation, see Appendix~\ref{app:stratified_proof}.

In terms of local unitary transformations, a generic pure three-qubit state requires five continuous parameters for its description. The state vector of a three-qubit pure state contains 16 real parameters; normalization and the global phase reduce this to 14, while the local action of $SU(2)^{\otimes3}$ removes another 9 degrees of freedom. Thus, even with $(x,y,z,t)$ fixed, one continuous local-unitary degree of freedom generally remains. Theorem~\ref{thm:five_body} provides the remaining fifth coordinate and establishes under which conditions it ceases to be an independent degree of freedom. Correspondingly, the five-dimensional compatibility region is fibered over $\Omega_4$, with the fibre dimension varying across strata.

We divide $\Omega_4$ into two parts:
\begin{equation}
\mathcal G = \{b\in\Omega_4: txyz>0,\ F_-<0\},
\label{eq:generic_stratum}
\end{equation}
and
\begin{equation}
\mathcal R = \Omega_4\setminus\mathcal G = \{b\in\Omega_4: txyz=0 \text{ or } F_-=0\}.
\label{eq:rigid_stratum}
\end{equation}
For points in $\mathcal G$, fixing $(x,y,z,t)$ still leaves $\mu$ free to vary continuously over a finite interval. Since different $\mu$ correspond to different Kempe invariants, these states are not locally unitarily equivalent to each other. That is, a point in $\mathcal G$ corresponds to a continuous family of pure states with the same $(x,y,z,t)$. In contrast, in $\mathcal R$, the allowed $\mu$ has only one value, so this extra continuous degree of freedom no longer exists.

The disappearance of the continuous Kempe degree of freedom corresponds to three distinct cases. First, when $t=0$, we have $q=p$, so from Eq.~\eqref{eq:mu_interval} we obtain
\begin{equation}
\mathcal F_b=\{p\}.
\label{eq:tzero_rigidity}
\end{equation}
This shows that in the region where the three-tangle vanishes, the reduced Kempe coordinate is uniquely determined.

Second, when $xyz=0$, we have $p=0$. From $\mu^2\le xyz$, it immediately follows that
\begin{equation}
\mu=0.
\label{eq:pairwise_degenerate_rigidity}
\end{equation}
Thus, whenever any one of the two-body concurrences is zero, the Kempe fibre degenerates to a point.

Third, on the nondegenerate boundary $F_-=0$ with $txyz>0$, Eq.~\eqref{eq:factor_relation_main} gives $q=t+p$, and hence
\begin{equation}
\mathcal F_b=\{p\}.
\label{eq:boundary_rigidity}
\end{equation}
Therefore, the fifth continuous coordinate is fixed on the boundary of the four-dimensional compatibility region.

From Eq.~\eqref{eq:G_mu_identity}, it follows that on this nondegenerate boundary we have $G(s)=0$.
When the coefficient of the phase term is nonzero, this condition is equivalent to $|\cos\phi|=1$,
so the phase in the Ac{\'i}n standard form can only take boundary values, meaning that there is no continuous degree of freedom on this boundary.

\section{Transfer to Other Entanglement Measures}
\label{sec:measure_transfer}

\subsection{Concurrence-generated and derived measures}

The first three coordinates can also be replaced by any bipartite entanglement measure $E=F(C)$ that is strictly increasing with concurrence. Define $\chi_E(e)=[F^{-1}(e)]^2$, then it suffices to replace $x, y, z$ respectively by
$\chi_E(E_{AB}),\chi_E(E_{AC}),\chi_E(E_{BC})$, to obtain the corresponding compatibility conditions.

Taking the entanglement of formation for two qubits as an example,
\begin{equation}
E_f(C) = h_2\!\left(\frac{1+\sqrt{1-C^2}}{2}\right),
\label{eq:eof_wootters}
\end{equation}
using the inverse relation of Wootters' formula, the five-dimensional compatibility condition can be directly written in terms of the three bipartite EOFs as coordinates~\cite{Wootters1998}.

The coordinates $(x,y,z,t)$ also fix several standard multipartite entanglement measures. For pure three-qubit states, the CKW relation can be written as
\begin{equation}
C_{A:BC}^{2}=x+y+t, C_{B:AC}^{2}=x+z+t, C_{C:AB}^{2}=y+z+t.
\label{eq:ckw_three}
\end{equation}
Therefore, the squared genuine multipartite concurrence~\cite{Ma2011GME} is
\begin{equation}
C_{\rm GME}^{2} = \min\{x+y+t,\ x+z+t,\ y+z+t\}.
\label{eq:GME_from_body}
\end{equation}
Similarly, the Meyer--Wallach multipartite entanglement measure~\cite{MeyerWallach2002} can also be directly expressed in terms of these four coordinates as
\begin{equation}
Q_{\rm MW} = t + \frac{2}{3}(x+y+z).
\label{eq:MW_from_body}
\end{equation}

The pairwise EOFs, the one-versus-rest concurrences, $C_{\rm GME}$, and $Q_{\rm MW}$ depend only on $(x,y,z,t)$ and are therefore constant along a Kempe fiber. Different points on a generic fiber can nevertheless be LU inequivalent. On the rigid strata of Sec.~\ref{sec:stratified_completeness}, the fifth coordinate is fixed.

\subsection{Entanglement-monotone parametrization of the fifth coordinate}
\label{subsec:monotone_fifth}

Although the Kempe invariants are local unitary invariants rather than entanglement monotones, they are algebraically the most convenient for describing the fifth continuous degree of freedom. This degree of freedom can also be equivalently characterized by entanglement monotones. For pure three-qubit states, Oreshkov and Brun constructed the following polynomial entanglement monotone~\cite{OreshkovBrun2006}:
\begin{equation}
\begin{aligned}
\phi_{ABC}
=&\ 69
-\operatorname{Tr}\!\left[\left(2\rho_{AB}+\rho_A\otimes I_B+I_A\otimes\rho_B\right)^3\right]\\
&-3\operatorname{Tr}\rho_{AB}^2 .
\end{aligned}
\label{eq:OB_monotone}
\end{equation}
For pure three-qubit states, $\phi_{ABC}$ is invariant under arbitrary permutations of the three qubits and is non-increasing on average under pure-state LOCC operations. Using
\begin{equation}
\operatorname{Tr}\rho_i^2 = 1-\frac12 C_{i:jk}^2, \operatorname{Tr}\rho_i^3 = 1-\frac34 C_{i:jk}^2,
\label{eq:qubit_trace_identities}
\end{equation}
together with the CKW relation in Eq.~\eqref{eq:ckw_three}, the polynomial given by Oreshkov and Brun can be reduced to
\begin{equation}
\phi_{ABC} = 12-12\kappa+27S+\frac{81}{2}t.
\label{eq:phi_kappa_relation}
\end{equation}
From Eq.~\eqref{eq:mu_def} we further obtain
\begin{equation}
\mu = 4S+\frac{11}{2}t-\frac{\phi_{ABC}}{9}.
\label{eq:mu_phi_relation}
\end{equation}
Therefore, when $(x,y,z,t)$ is fixed, $\phi_{ABC}$, $\kappa$, and $\mu$ are merely different parametrizations of the same fifth continuous degree of freedom. That is, the fifth continuous coordinate can be expressed either in terms of the Kempe invariants or equivalently in terms of the Oreshkov--Brun entanglement monotones.

\begin{corollary}[Entanglement-monotone fifth coordinate]
\label{cor:monotone_fifth}
Taking the Oreshkov--Brun monotone $\phi_{ABC}$ as the fifth coordinate, the five-dimensional compatibility region can be equivalently represented in the coordinates $(x,y,z,t,\phi_{ABC})$. Given a set of coordinates $(x,y,z,t,\phi_{ABC})$, it corresponds to some normalized pure three-qubit state if and only if Eq.~\eqref{eq:five_base_condition} holds and, in addition, the following inequalities are satisfied:
\begin{equation}
\left( 4S+\frac{11}{2}t-\frac{\phi_{ABC}}{9} \right)^2 \le xyz,
\label{eq:phi_condition_1}
\end{equation}
and
\begin{equation}
4S+\frac{13}{2}t-\frac{\phi_{ABC}}{9}
\ge \sqrt{(t+x)(t+y)(t+z)}.
\label{eq:phi_condition_2}
\end{equation}
\end{corollary}

When $(x,y,z,t)$ is fixed, Eq.~\eqref{eq:phi_kappa_relation} gives an affine relation with nonzero slope between $\phi_{ABC}$ and $\kappa$. Using $\phi_{ABC}$ instead of the Kempe invariant does not change the
fiber structure.  In the generic region the fifth coordinate remains
continuous, while on the rigid strata it is fixed.  The Kempe invariant
is kept in the main theorem because the resulting compatibility
conditions are algebraically simpler. The Oreshkov--Brun monotone describes the same fifth degree of freedom, while the three pairwise concurrences retain their usual meaning as bipartite entanglement measures.

\section{Discussion and Conclusion}

The main result of this paper is that $(x,y,z,t)$ do not completely determine the local-unitary structure of a pure three-qubit state. Generically, one continuous degree of freedom remains after these four coordinates are fixed. This freedom is lost on certain lower-dimensional strata. For non-degenerate boundaries, this change simultaneously corresponds to the saturation of the Ac\'{\i}n phase condition, and the difference between the boundary and the interior region can also be seen directly from the standard form.

Quantities fixed by $(x,y,z,t)$ are constant along a Kempe fiber and do not capture the remaining continuous LU degree of freedom. States on the same fiber can have the same standard tangle-based entanglement measures while belonging to different LU classes. We use the Kempe invariant because it gives the simplest algebraic form for this fifth coordinate. The same degree of freedom can instead be parametrized by the Oreshkov--Brun pure-state entanglement monotone.

A question still worth further investigation is whether states at different points on the same Kempe fiber can behave differently in concrete quantum-information tasks despite sharing the same $(x,y,z,t)$. There has already been some research on the operational meaning of the Kempe invariant and its changes under local operations; however, if $(x,y,z,t)$ are fixed and only the fifth continuous coordinate is varied, there is still no clear answer as to what physical difference such a change corresponds to.

\acknowledgments
This work was supported by the Natural Science Foundation of Anhui Province under Grant No. 2508085ZD001, and Scientific Research Projects of Anhui Provincial Department of Education under Grant Nos. 2025AHGXZK20126 and 2025AHGXZK50054.

\appendix

\section{Ac{\'i}n Feasibility and the Reduced Kempe Coordinate}
\label{app:kempe_acin}

Using the Ac{\'i}n standard form, any pure three-qubit state can be written, up to local unitary transformations, as~\cite{Acin2000}
\begin{equation}
|\psi\rangle = a|000\rangle + b e^{i\phi}|100\rangle + c|101\rangle + d|110\rangle + e|111\rangle,
\label{eq:acin_main}
\end{equation}
where $a,b,c,d,e\ge0$, $\phi\in[0,\pi]$, and the amplitudes satisfy the normalization condition. Let $s=4a^2$.
When $s>0$, the four tangle coordinates directly determine
\begin{equation}
c^2=\frac{y}{s}, d^2=\frac{x}{s}, e^2=\frac{t}{s},
\label{eq:acin_tangles}
\end{equation}
and from the normalization condition we obtain
\begin{equation}
b^2=\frac{K(s)}{s}, \qquad K(s)=s-\frac{s^2}{4}-x-y-t.
\label{eq:Kmain5}
\end{equation}
The $BC$ concurrence then gives
\begin{equation}
\frac{s^2z}{4} = xy+tK(s)-2\sqrt{xytK(s)}\cos\phi.
\label{eq:zrelation_supp5}
\end{equation}
Under the condition $K(s)\ge0$, an admissible real phase $\phi$ exists if and only if
\begin{equation}
G(s) = \left( xy+tK(s)-\frac{s^2z}{4} \right)^2 -4xytK(s) \le0.
\label{eq:Gfive_supp}
\end{equation}
If the coefficient of $\cos\phi$ is zero, then this concurrence condition itself no longer depends on the phase $\phi$.

In the Ac{\'i}n standard form, the Kempe invariant can be calculated as
\begin{equation}
\kappa(s) = 1-\frac32t-\frac34S +\frac{3}{16}(t+z)s +\frac{3}{4s}(t+x)(t+y).
\label{eq:kappas_main}
\end{equation}
Substituting the definition of the reduced coordinate $\mu$, we have
\begin{equation}
\mu(s) = -t+\frac{t+z}{4}s +\frac{(t+x)(t+y)}{s}.
\label{eq:mu_s}
\end{equation}
On the other hand, the phase constraint polynomial satisfies
\begin{equation}
\frac{G(s)}{s^2} = \mu(s)^2-xyz.
\label{eq:G_mu_identity}
\end{equation}
Therefore, the phase realizability condition $G(s)\le0$ in the Ac{\'i}n standard form corresponds exactly to $\mu^2\le xyz$.
When $t+z>0$, from Eq.~\eqref{eq:mu_s}, using the arithmetic--geometric mean inequality, we obtain
\begin{equation}
\begin{aligned}
\mu+t &= \frac{t+z}{4}s +\frac{(t+x)(t+y)}{s}\\
&\ge \sqrt{(t+x)(t+y)(t+z)} =q.
\end{aligned}
\label{eq:mu_AMGM}
\end{equation}
Thus, the realizability condition in the Ac{\'i}n parameters is directly translated into the two constraints satisfied by the fifth coordinate in Theorem~\ref{thm:five_body}.

\section{Proof of Theorem~\ref{thm:five_body}}
\label{app:five_body_proof}
The four-dimensional part has already been given by the exact realizability result of Ref.~\cite{Allen2017}; here it is merely re-expressed using the notation of squared concurrence. For each point of $\Omega_4$, Eq.~\eqref{eq:five_mu_condition} then provides the necessary and sufficient condition on the fifth coordinate.

\paragraph{Necessity.}
Take first an Ac{\'i}n representative with $s>0$. From
Eq.~\eqref{eq:G_mu_identity} and $G(s)\le0$, we have
\begin{equation}
\mu^2\le xyz.
\end{equation}
For $t+z>0$, Eq.~\eqref{eq:mu_AMGM} gives the second condition in
Eq.~\eqref{eq:five_mu_condition}.

If $t+z=0$, then $t=z=0$. The concurrence relation gives $xy=0$,
and Eq.~\eqref{eq:G_mu_identity} gives $\mu=0$. Since $q=0$, the
second condition is saturated.

For $s=0$, we have $a=0$, and the Ac{\'i}n state factorizes as
$|1\rangle_A\otimes|\chi\rangle_{BC}$. Hence $x=y=t=0$, while
$z$ may remain nonzero. For this product form,
$\kappa=1-3z/4$, and Eq.~\eqref{eq:mu_def} gives $\mu=0$. Both
conditions in Eq.~\eqref{eq:five_mu_condition} are therefore satisfied.

\paragraph{Sufficiency for $txyz>0$.}
Let $(x,y,z,t)\in\Omega_4$ with $txyz>0$, and choose $\kappa$
satisfying Eq.~\eqref{eq:five_mu_condition}. Equation~\eqref{eq:mu_s}
gives
\begin{equation}
\frac{t+z}{4}s^2
-(\mu+t)s
+(t+x)(t+y)=0.
\label{eq:s_quadratic_mu}
\end{equation}
The second condition in Eq.~\eqref{eq:five_mu_condition} gives
\begin{equation}
(\mu+t)^2
\ge
q^2
=
(t+x)(t+y)(t+z),
\end{equation}
so the two roots are real,
\begin{equation}
s_\pm
=
\frac{2\left[
\mu+t\pm\sqrt{(\mu+t)^2-q^2}
\right]}{t+z}.
\label{eq:sroots_mu}
\end{equation}
Their sum and product are positive, hence $s_\pm>0$.

For either root, Eq.~\eqref{eq:G_mu_identity} gives
\begin{equation}
G(s)=s^2(\mu^2-xyz)\le0.
\end{equation}
Since $xyt>0$, $K(s)<0$ would make both terms in
Eq.~\eqref{eq:Gfive_supp} positive, in contradiction with $G(s)\le0$.
Thus $K(s)\ge0$. Equations~\eqref{eq:acin_tangles} and
\eqref{eq:Kmain5} then define real nonnegative amplitudes satisfying
the normalization condition, while $G(s)\le0$ guarantees an admissible
phase through Eq.~\eqref{eq:zrelation_supp5}. The resulting Ac\'{\i}n
state has the prescribed $x,y,z,t$ and $\mu$, and therefore the required
Kempe invariant by Eq.~\eqref{eq:mu_def}.

\paragraph{Degenerate sectors.}
For $t=0$, one has $q=p$. Equation~\eqref{eq:five_mu_condition}
then gives $-p\le\mu\le p$ and $\mu\ge p$, so
\begin{equation}
\mu=p.
\label{eq:tzero_mu}
\end{equation}
The four-dimensional achievability result of Ref.~\cite{Allen2017} provides
a pure-state realization of every point $(x,y,z,0)\in\Omega_4$, and
the necessity argument fixes its fifth coordinate to this value.

For $t>0$ and $xyz=0$, we have $p=0$, and the first condition in
Eq.~\eqref{eq:five_mu_condition} gives $\mu=0$. On $\Omega_4$,
\begin{equation}
(t+p)^2-q^2=-tF_-,
\label{eq:factor_relation_supp}
\end{equation}
with $F_-\le0$. Hence $t\ge q$, and the second condition is satisfied
as well. Ref.~\cite{Allen2017} supplies a pure-state realization of the
base point, whose fifth coordinate is fixed to $\mu=0$ by necessity.

The factorized edge $s=0$, where $x=y=t=0$, belongs to the $t=0$
sector above.\hfill\(\square\)

\section{Explicit Reconstruction and Degenerate Representatives}
\label{app:constructive}

For $txyz>0$, either root in Eq.~\eqref{eq:sroots_mu} may be used in
the Ac\'{\i}n parametrization. The amplitudes are
\begin{equation}
\begin{aligned}
a&=\frac{\sqrt{s}}{2},&
b&=\sqrt{\frac{K(s)}{s}},&
c&=\sqrt{\frac{y}{s}},\\
d&=\sqrt{\frac{x}{s}},&
e&=\sqrt{\frac{t}{s}}.&&
\end{aligned}
\label{eq:construct5}
\end{equation}
For $K(s)>0$, the phase is fixed by
\begin{equation}
\cos\phi
=
\frac{xy+tK(s)-s^2z/4}
{2\sqrt{xytK(s)}}.
\label{eq:phase_reconstruct5}
\end{equation}
The inequality $G(s)\le0$ ensures $|\cos\phi|\le1$. If $K(s)=0$,
the phase-dependent term in Eq.~\eqref{eq:zrelation_supp5} vanishes
and the concurrence condition is independent of $\phi$.
Substitution into Eqs.~\eqref{eq:acin_tangles},
\eqref{eq:zrelation_supp5}, and \eqref{eq:mu_s} recovers the prescribed
coordinates.

For $t=0$ and $p>0$, Eq.~\eqref{eq:tzero_mu} gives $\mu=p$, and
Eq.~\eqref{eq:s_quadratic_mu} has the double root
\begin{equation}
s=2\sqrt{\frac{xy}{z}}.
\label{eq:tzero_s}
\end{equation}
At this value,
\begin{equation}
zK(s)=2p-Q=-F_-(x,y,z,0)\ge0,
\end{equation}
so $K(s)\ge0$ and Eq.~\eqref{eq:construct5} gives real amplitudes.
Since $t=0$, Eq.~\eqref{eq:zrelation_supp5} is independent of the phase.

For $t>0$ and $xyz=0$, the degenerate-sector argument above fixes
$\mu=0$. Substitution into Eq.~\eqref{eq:s_quadratic_mu} gives the
candidate values of $s$. The physical realization established in the
proof above corresponds to a positive root with $K(s)\ge0$. For such a
root the amplitudes are again given by Eq.~\eqref{eq:construct5}. When
$xytK(s)>0$, the phase follows from Eq.~\eqref{eq:phase_reconstruct5};
if this coefficient vanishes, Eq.~\eqref{eq:zrelation_supp5} contains
no phase dependence.

The remaining $t=0$, $p=0$ cases are the single-pair edges. Indeed,
for $t=0$ and $p=0$ the base condition gives $F_-=Q\le0$, while
$Q=xy+xz+yz\ge0$. Hence $Q=0$, so at most one of $x,y,z$ can be
nonzero. For $r\in[0,1]$, define
\begin{equation}
\alpha_r
=
\sqrt{\frac{1+\sqrt{1-r}}{2}},
\qquad
\beta_r
=
\sqrt{\frac{1-\sqrt{1-r}}{2}},
\end{equation}
so that
\begin{equation}
4\alpha_r^2\beta_r^2=r.
\end{equation}
The states
\begin{align}
|\psi_z^{(BC)}\rangle
&=
|1\rangle_A\otimes
\left(
\alpha_z|00\rangle+\beta_z|11\rangle
\right)_{BC},\\
|\psi_x^{(AB)}\rangle
&=
\alpha_x|000\rangle+\beta_x|110\rangle,\\
|\psi_y^{(AC)}\rangle
&=
\alpha_y|000\rangle+\beta_y|101\rangle
\end{align}
have, respectively, $z$, $x$, and $y$ as their only nonzero squared
pairwise concurrence. The fully separable origin is represented by
$|000\rangle$.

\section{Proof of the Stratified-Completeness Corollary}
\label{app:stratified_proof}

For fixed $b=(x,y,z,t)\in\Omega_4$, the reduced Kempe coordinate
ranges over
\begin{equation}
\mathcal F_b
=
[\max\{-p,q-t\},p].
\end{equation}
This interval has nonzero length if and only if
\begin{equation}
p>\max\{-p,q-t\},
\end{equation}
or, equivalently,
\begin{equation}
p>0,
\qquad
t+p>q.
\end{equation}
Since $p=\sqrt{xyz}$, the first inequality is $xyz>0$.
For $t=0$, $q=p$, so the second inequality is saturated. For $t>0$,
Eq.~\eqref{eq:factor_relation_supp} gives
\begin{equation}
t+p>q
\quad\Longleftrightarrow\quad
F_-<0.
\end{equation}
Therefore
\begin{equation}
\dim_{\rm cont}\mathcal F_b=1
\quad\Longleftrightarrow\quad
txyz>0\ \text{and}\ F_-<0.
\end{equation}
Otherwise the fiber is a single point, which gives
Eq.~\eqref{eq:fiber_dimension}.

For $t=0$, $q=p$, and
\begin{equation}
\mathcal F_b=\{p\}.
\end{equation}
For \(xyz=0\), the inequality $\mu^2\le xyz$ gives
\begin{equation}
\mu=0.
\end{equation}
On the boundary $F_-=0$ with $txyz>0$,
Eq.~\eqref{eq:factor_relation_supp} gives $q=t+p$, and hence
\begin{equation}
\mathcal F_b=\{p\}.
\end{equation}
In this case $\mu^2=xyz$, so Eq.~\eqref{eq:G_mu_identity} gives
\begin{equation}
G(s)=0.
\end{equation} \hfill\(\square\)

\end{document}